\documentclass[aps,pra,twocolumn,superscriptaddress,longbibliography]{revtex4-1}

\usepackage{xcolor}
\usepackage[colorlinks=true, linkcolor=blue, citecolor=blue, urlcolor=blue]{hyperref}
\usepackage{amsmath, amssymb, amsthm, float, graphicx,bm}
\usepackage{amsfonts,braket}
\usepackage{graphics,float}
\usepackage{subfigure}
\usepackage{todonotes}
\usepackage[left=1.7cm,right=1.7cm,top=2cm, bottom=2cm]{geometry}
\usepackage{appendix}

\usepackage{soul}

\begin{document}
	\title{Near-100\% two-photon-like coincidence-visibility dip with classical\\light and the role of complementarity}
	
	\author{Simanraj Sadana}
	\affiliation{Raman Research Institute, Sadashivanagar, Bangalore 560 080, India}
	\author{Debadrita Ghosh}
	\affiliation{Raman Research Institute, Sadashivanagar, Bangalore 560 080, India}
	\author{Kaushik Joarder}
	\affiliation{Raman Research Institute, Sadashivanagar, Bangalore 560 080, India}
	\author{A.~Naga Lakshmi}
	\affiliation{Raman Research Institute, Sadashivanagar, Bangalore 560 080, India}
	\author{Barry C. Sanders}
	\affiliation{Raman Research Institute, Sadashivanagar, Bangalore 560 080, India}
	\affiliation{Institute for Quantum Science and Technology, University of Calgary, Calgary, Alberta, Canada T2N 1N4}
	\affiliation{Program in Quantum Information Science, Canadian Institute for Advanced Research, Toronto, Ontario, Canada M5G 1Z8}
	\author{Urbasi Sinha}
	\email{usinha@rri.res.in}
	\affiliation{Raman Research Institute, Sadashivanagar, Bangalore 560 080, India}

	\begin{abstract}
		The Hong-Ou-Mandel effect is considered a signature of the quantumness of light, as the dip in  coincidence probability using semiclassical theories has an upper bound of 50\%. Here we show, theoretically and experimentally, that,
		with proper phase control of the signals, classical pulses can mimic a Hong-Ou-Mandel-like dip. We demonstrate a dip of $(99.635 \pm 0.002)\%$ with classical microwave fields.
		Quantumness manifests in wave-particle complementarity of the two-photon state. We construct quantum and classical interferometers for the complementarity test and show that while the two-photon state shows wave-particle complementarity the classical pulses do not.
	\end{abstract}
	
	\maketitle
	
	\section{Introduction}
	The Hong-Ou-Mandel (HOM) two-photon coincidence-visibility dip (TPCVD)~\cite{HOM87,FL87}, is a salient fourth-order interference effect~\cite{HS96}, {and} one of the most important effects and tools in quantum optics. Myriad applications include measuring photon purity~\cite{CLS10} and distinguishability~\cite{Ou07}, heralding in optical quantum computing gates~\cite{KLM01},
	conceptually underpinning the complexity
	of the boson sampling problem~\cite{AA13}, and realizing a NOON state for quantum metrology~\cite{Dow08}.
	
	TPCVD is manifested by injecting a single photon into each input port of a balanced beam splitter and observing anticorrelated output in the form of bunching. Mathematically, the input is the pure product state~$\ket{11}$ {and} the output is the symmetric or antisymmetric superposition of $\ket{20}$ and $\ket{02}$, resulting in a coincidence probability of zero. 
	
	Given some \text{delay} $\tau$ between photon {arrivals} at the two input ports, the coincidence probability $C(\tau)$ drops as $\tau$ approaches zero. 
	The TPCVD,
	\begin{align}
	V:=\frac{C(\infty)-C(0)}{C(\infty)}=1-\frac{C(0)}{C(\infty)},
	\label{eq:visibilityDef}
	\end{align}
	is unity in the ideal case~\cite{KSC92, Ish2016}.
	
	Despite the immense importance, applicability, and success of HOM TPCVD,
	a widespread misconception {is} that exceeding a 50\% TPCVD falsifies classical electromagnetic field theory.
	Example quotations include 
	``fourth-order interference of classical fields cannot yield visibility larger than 50\%'' \cite{Ou90},
	``as long as the visibility of the coincidence dip is greater than 50\%,
	no semiclassical field theory can account
	for the observed interference'' \cite{KSC92},
	``visibility, being greater than 50\%,
	is clear evidence of non-classical interference'' \cite{RTL05}, and
	``classical theory of the coherent superposition of electromagnetic waves, however, can only explain a HOM dip with V$\leq$0.5'' \cite{KSKT13}.
	This myth matters as the 50\% dip threshold is widely accepted as proving that two-photon interferometry has entered the quantum domain.
	
	The 50\% TPCVD is believed to be the threshold between classical and quantum behavior of light, because independent classical pulses with uniformly randomized relative phase yield $V=1/2$~\cite{Car,KSKT13}.
	{The argument}
	against the classical description is that it fails to predict the observed 100\% visibility and, instead,
	puts an upper bound of 50\% on it. 
	
	In this paper, we perform two precision experiments and show that if the phase between the input signals is randomized over a preselected set, 100\% visibility can be achieved even with classical pulses. Therefore, the aforementioned
	reason for the failure of the classical description of light is invalid. The real quantum signature lies in completing the interferometer as a two-photon analog of the experimental proof of complementarity for a single photon~\cite{GRA86}. Curiously, {this subtle yet important point has no experimental demonstration in the literature.}
	
	\section{Theory}
	In the semiclassical theory of photodetection~\cite{Mandel1964,Mandel1958}, the coincidence probability at the two detectors is proportional to the cross correlation of the integrated intensities at the detectors. The normalized correlation function is
\begin{widetext}
	\begin{equation}
		\label{eq:crossCorrelationTheoryExpression}
		C(\tau)
		:=\frac{\int d\varphi P\left(\varphi\right)
			\int_{T_\text{on}}^{T_\text{off}} dt\left|E_+(t;\omega,\tau,\varphi)\right|^2
			\int_{T_\text{on}}^{T_\text{off}} dt'\left|E_-(t';\omega,\tau,\varphi)\right|^2}
		{\left[\int d\varphi P\left(\varphi\right)
			\int_{T_\text{on}}^{T_\text{off}} dt
			\left|E_+(t;\omega,\tau,\varphi)\right|^2
			\right]
			\left[\int d\varphi'P\left(\varphi'\right)
			\int_{T_\text{on}}^{T_\text{off}} dt
			\left|E_-(t;\omega,\tau,\varphi')\right|^2
			\right]},
	\end{equation}
\end{widetext}
	where $E_+(t;\omega, \tau, \varphi)$, $E_-(t;\omega, \tau, \varphi)$ are the output fields from the transmitted and reflected ports of the 50:50 {beam} splitter respectively, when the input pulses have carrier frequency $\omega$, a time delay $\tau$ between them, and relative phase $\varphi$ which fluctuates with probability $P(\varphi)$.

	If $T_\text{off} - T_\text{on}$ is large compared to input-pulse width, Eqs.~(\ref{eq:visibilityDef}) and (\ref{eq:crossCorrelationTheoryExpression}) yield
	\begin{equation}
	V = \int  d\varphi~ P(\varphi)\cos^2\varphi,
	\label{eq:tpcvdWithPhi}
	\end{equation}
	which depends on~$P(\varphi)$,
	and detailed calculations are in Appendix \ref{append:TPCVDSemiclassicalTheory}.
	Uniformly randomized~$\varphi\in\left[0,2\pi\right)$
	is equivalent to no control over $\varphi$
	so $V = 1/2$,
	which is widely accepted as the quantum-classical boundary, whereas we show that a $50\%$ dip is a consequence of ignorance or lack of phase control.
	
	Choosing different probability distributions for $\varphi$ yields different TPCVDs.
	For example, $V=0$ if $\varphi\in\left\{\frac{\pi}{2},\frac{3\pi}{2}\right\}$ and $V=1/2$ if $\varphi\in\left\{0,\frac{\pi}{2},\pi,\frac{3\pi}{2}\right\}$, with uniform distribution over these respective sets.
	Significantly, $V=1$, i.e., TPCVD is 100\%, for uniformly randomized $\varphi \in \{0,\pi\}$.
	
	If exceeding the 50\% dip threshold does not signify quantumness,
	where does quantumness appear?
	\emph{Contraria sunt complementa},
	or ``opposites are complementary,'' as Niels Bohr's coat of arms says:
	a particle is opposite to a wave.
	In the context of complementarity,
	particlelike behavior is characterized by detection
	that reveals indivisibility of the object.
	Wavelike behavior,
	on the other hand,
	is characterized by interference fringes that signify the object is delocalized and coherent across some region.
	For instance, single-photon complementarity is demonstrated by a particlelike test that shows that a photon passing through a beam splitter is indivisible.
	The wavelike property is demonstrated
	by allowing this indivisible entity to take two different paths
	and observing interference fringes as a function of path-length difference through a difference measurement by the photodetectors at the two output ports~\cite{GRA86}.
	
	We extend the above argument to the TPCVD experiment. In the quantum version of the experiment two indistinguishable photons enter a balanced beam splitter, one in each port. The output is a superposition of a two-photon state in each arm, with $100\%$ TPCVD showing particlelike behavior. Recombining the two arms using a second beam splitter yields an interference between them, confirming that output of the first beam splitter is in quantum superposition, a signature of wavelike behavior. Thus the two-photon state shows wave-particle complementarity.
	
	In the classical version, 
	inputs are classical pulses with $\varphi \in \{0,\pi\}$ with equal probability, resulting in $100\%$ TPCVD showing a particlelike behavior. However, unlike the quantum case, both pulses exit the beam splitter either from one port or the other without superposition of paths, thus retaining its particlelike behavior. The lack of superposition is confirmed when no interference is observed on recombining the two paths using a second beam splitter. Therefore, the classical pulse does not show wave-particle complementarity.
	Thus, quantumness arises through
	two-photon complementarity, not a TPCVD of more than $50\%$. We provide a mathematical description of this phenomenon in Appendix \ref{app:compTest}.
	
	\section{Classical experiment}
\begin{figure}
	\includegraphics[width=0.48\textwidth]{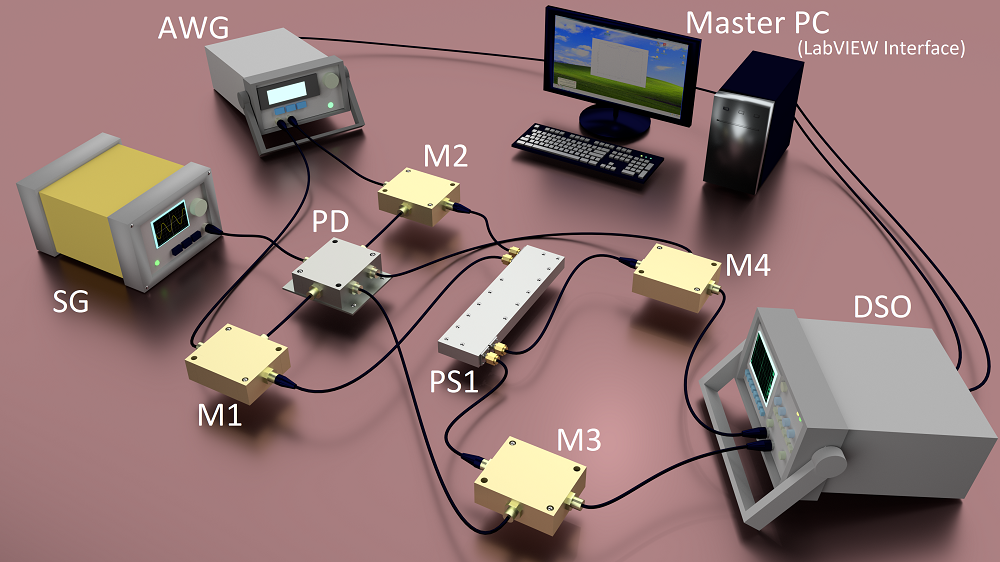}
	\caption{\label{fig:schematicsC}Schematic for the HOM experiment using classical microwave pulses. Detailed description of the setup is in the main text.
	}
\end{figure}

	The key to demonstrating 100\% TPCVD with classical pulses is controlling $\varphi$,
	which is achieved in an optics experiment using electro-optic modulators (EOMs)~\cite{Gobert11,Feng13,hergott2011carrier}.
	However, the TPCVD accuracy depends on the input pulses being identical.
	Pulse distortion caused by EOMs can increase distinguishability,
	thereby reducing the TPCVD.
	Moreover, maintaining identical polarizations, perfect alignment of the beams, as well as active phase control in such an experiment that requires many iterations, is cumbersome.
	This challenge motivated our use of a carefully designed electrical version of the beam splitter experiment shown in Fig.~\ref{fig:schematicsC}.
	As electrical signals with accurate phase control are directly generated,
	an extra apparatus for phase control is not needed.
	This innovation also eliminates polarization and alignment issues leading to a clean, precise experiment beyond the capability of optical systems,
	which enabled tremendous accuracy and precision in demonstrating 100\% TPCVD with classical fields.
	
	An arbitrary wave-form generator (AWG) (Keysight Technologies-33622A) with dual channels is programed using \textsc{\small LABVIEW} to generate two sine signals with Gaussian amplitude modulation with 100-mV peak-to-peak and 1-KHz carrier frequency. The programmability of AWG facilitates the control of the relative phase $\varphi$ and the time delay $\tau$ between the two pulses. The input pulses are injected into a 180$^\circ$ power splitter PS1.
	
	The two power splitters PS1 and PS2 are broadband power splitters (ET Industries model J-076-180) with operating frequency from 770~MHz to 6~GHz. Both power splitters are characterized, and from $S$-parameter analysis~\cite{pozar2009microwave}, the operational frequency is chosen to be 1162~MHz, {for which} the power splitting ratios {are closest to the ideal expectation} of 50:50 (49.3:50.7 for PS1 and 50.3:49.7 for PS2) and phase error is around~2$-$3$^\circ$.
	
	{An} oscilloscope (Agilent Technologies DSO60 14A) is used as detector with frequency detection up to 100~MHz. As the desired operation frequency for the power splitters is 1162~MHz, signals from AWG {are} up-converted at the transmitter end and the down-converted at receiver end which requires mixers and a local oscillator. The {local oscillator (SG)} (Keysight Analog Signal generator N5173B), {is} tuned at 1161.999~MHz {and} connected to a one-input-four-output power divider (Mini circuits part number: ZA4PD-2). Mixers M1, M2, M3 and M4 (Mini-circuits FM-2000, level 7 mixer), which operate in the range of 100-2000~MHz, are driven {by} four output ports of the power divider. The power level from {SG} is set to +14~dBm to take care of power loss in the cable and power divider. The output shown on the oscilloscope screen is saved as a \textsc{\small MATLAB}\textsuperscript{\tiny\textregistered} data file {for postprocessing}. Detailed calculations of the signal processing are shown in Appendixes \ref{append:signalProc} and \ref{append:dataAcPostProc}.

	{Using \textsc{\small LABVIEW} we vary $\tau$ between $-7$ and $7$ ms in steps of $1$ ms to calculate the cross correlation of the output signals for different time delays. 
	The output signals recorded at the two channels of oscilloscope are postprocessed to calculate the cross correlation using Eq.~(\ref{eq:crossCorrelationTheoryExpression}) and produce the plots of the cross correlation as a function $\tau$. The number of samples of pairs of input signal produced for good statistics is explained in Appendix \ref{sec: number of samples}.
	
	For $\varphi\in\{0,\pi\}$ with equal probability, a TPCVD of $\left(99.635 \pm 0.002\right)\%$ is achieved, as shown in Fig.~\ref{fig:classical100}. For $\varphi\in\{0,\pi/2,3\pi/2,\pi\}$,
	the TPCVD is $\left(48.03 \pm 0.035\right)\%$ shown in Fig.~\ref{fig:classical50}. We show the 50\% TPCVD for $\varphi \in [0,2\pi)$ in Appendix \ref{append:50continuous}.
	\begin{figure}
		\centering
		\includegraphics[width=0.49\textwidth]{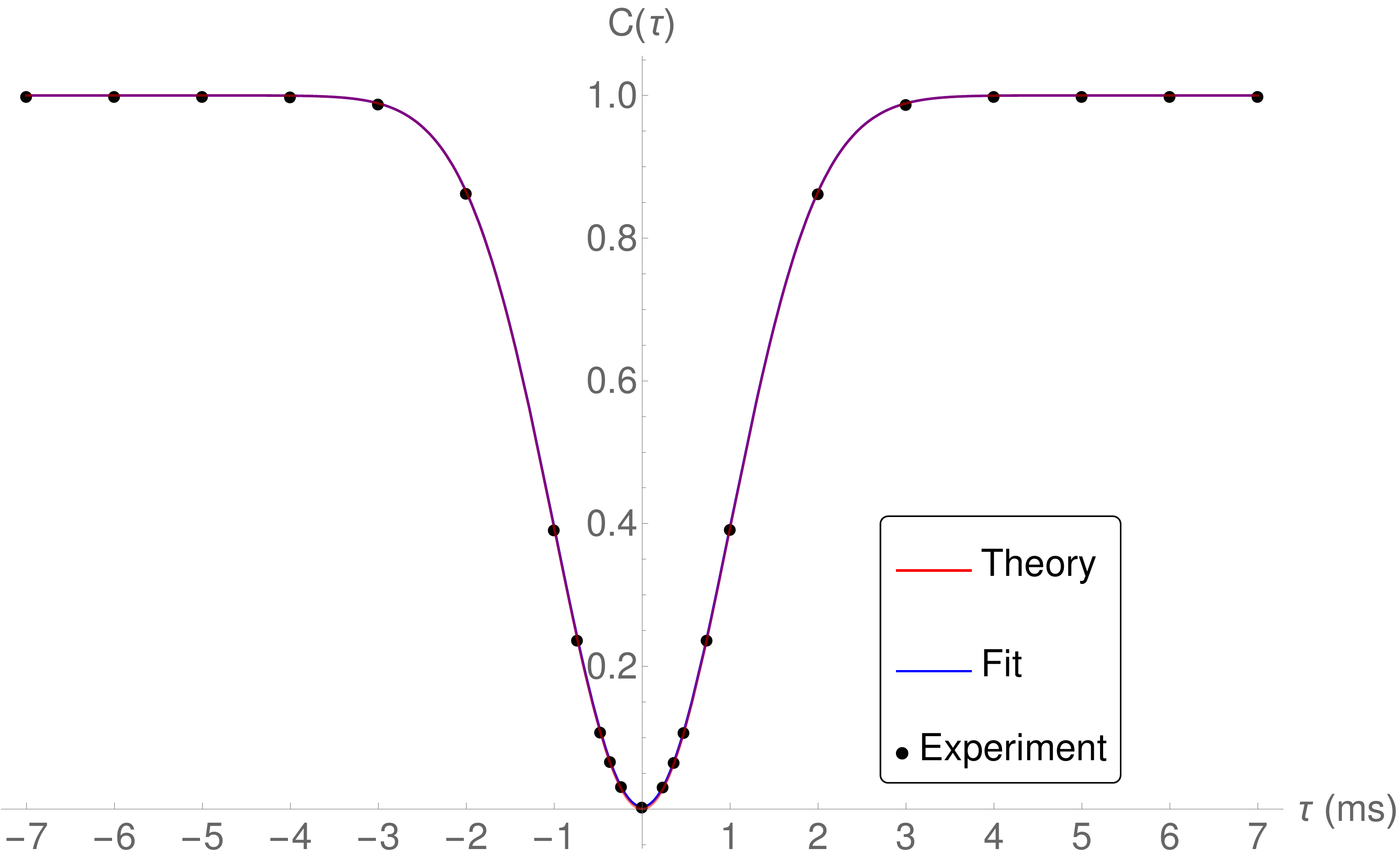}
		\caption{The plot shows the normalized cross correlation as a function  of time delay~$\tau$ when the phase~$\varphi$ between the two input pulses to the beam splitter is averaged over the set $\{0,\pi\}$ with choice of $0$ and $\pi$ being equally likely. The dots represent the experimental result, the green dashed line is the theoretical expectation, and the orange solid line is the result of fitting the theoretical expression for the cross correlation to the experimental result, with amplitude mismatch between the input pulses as the fitting parameter. The fitted curve overlaps almost completely with the theoretically expected curve.
			We obtain a TPCVD of $99.635\%$ with the error bars representing $95\%$ confidence interval of the mean value at each time delay.}
		\label{fig:classical100}
	\end{figure}
	\begin{figure}
		\centering
		\includegraphics[width=0.49\textwidth]{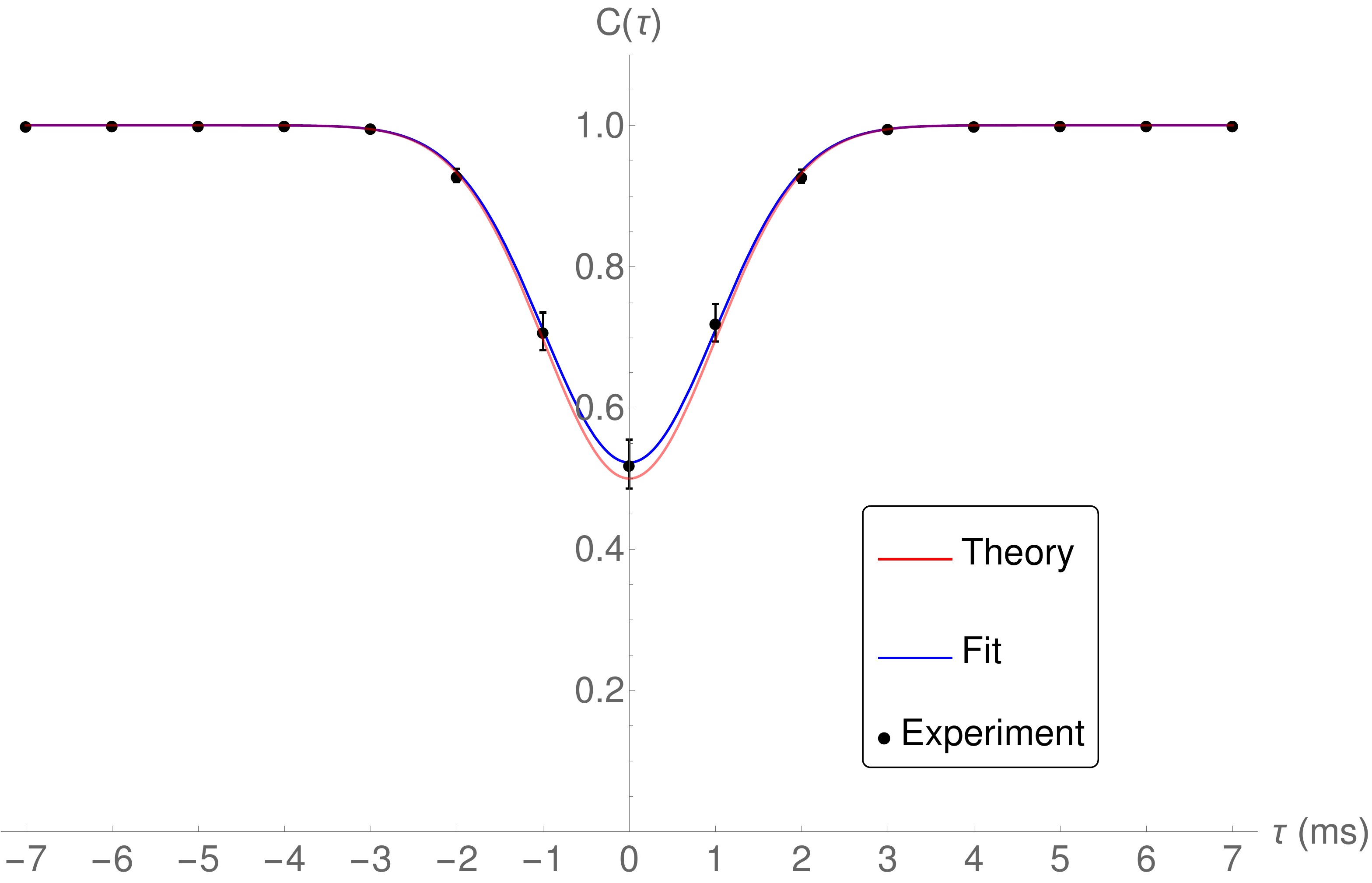}
		\caption{The plot shows the normalized cross correlation as a function  of time delay~$\tau$ when the phase~$\varphi$ between the two input pulses is averaged over the set $\{0,\pi/2, \pi, 3\pi/2\}$ with all four values of phase difference being equally likely. The dots represent the experimental result, the green dashed line is the theoretical expectation, and the orange solid line is the result of fitting the theoretical expression for the cross correlation to the experimental result, with amplitude mismatch between the input pulses as the fitting parameter. We obtain a TPCVD of $48.03\%$ with the error bars representing $95\%$ confidence interval of the mean value at each time delay.}
		\label{fig:classical50}
	\end{figure}
	
	\section{Quantum experiment}
	In the {quantum version} (see Fig.~\ref{fig:schematicsQ} for setup), we employ a type-II heralded photon source~\cite{Rubin1994} in {collinear} configuration. A diode laser (Toptica-iwave) at 405~nm with 50-mW power pumps a 5~mm $\times$ 5~mm $\times$ 10~mm type-II BBO crystal. A pair of lenses of focal lengths 22.5 and 25~cm focus the pump beam at the crystal and collimate the down-converted photons, respectively. Pairs of orthogonally polarized frequency degenerate photons with central wavelength of 810~nm are split in two directions by a polarizing beam splitter (PBS). A half wave plate in one of the arms of the PBS makes the two photons have identical polarization. An arrangement of long pass filters {transmits wavelengths above 405 nm}. A band-pass filter with a 3.1~nm bandwidth centered at 810~nm restricts the bandwidth of the transmitted photons also blocking any residual pump. Two fiber couplers collect photons from the same pair and inject them in a 2$\times$2 polarization-maintaining fused fiber beam splitter (FBS).
	
	\begin{figure}
		\includegraphics[width=0.48\textwidth]{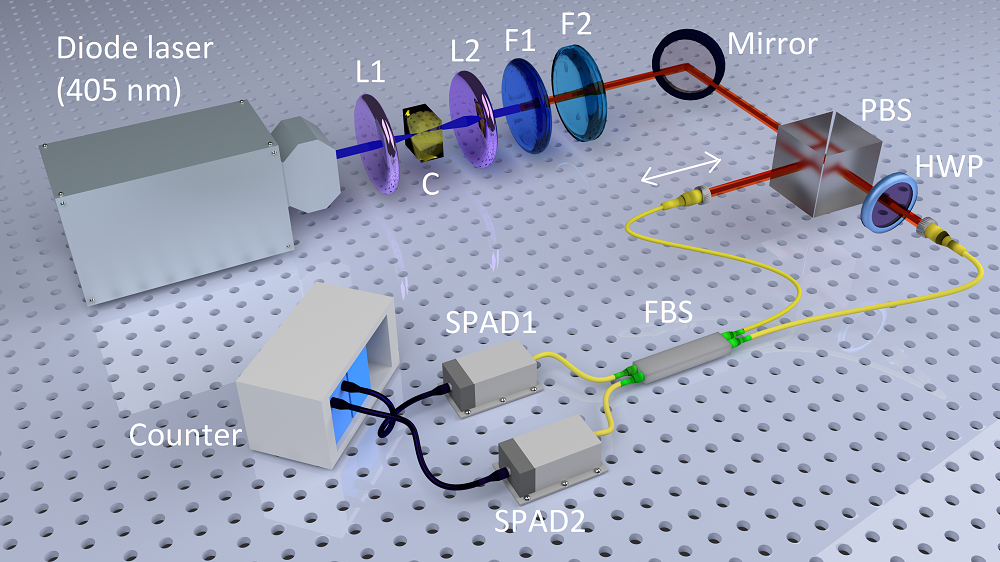}
		\caption{%
			Schematic for the HOM experiment using IR photons. Detailed description of the setup is in the main text.
		}
		\label{fig:schematicsQ}
	\end{figure}

	One of the couplers {is} mounted on a motorized stage {to vary} the delay between the two photons. We measure the coincidence count as a function of relative beam displacement between the signal and idler photons.
	A 10-$\mu$m step size for the stage provides necessary precision to resolve the HOM-dip width.
	The data acquisition time is 5~s at each position and we measure at {160} positions. We repeat the measurement 100 times for better averaging and to estimate an error bar (see Appendix \ref{append:bootstrapPhoton} for details).
	
	Figure~\ref{fig:IR} shows two-photon coincidence counts as a function of time delay~$\tau$ between the two input photons. We obtain a TPCVD of $\left(96.06 \pm 0.16\right)\%$. {Details} concerning choice of parameters for both experimental setups are in Appendixes \ref{app:error}, \ref{app:ThEstQ}, and \ref{app:ThFitQ}. 
	
	\begin{figure}
		\centering
		\includegraphics[width=0.49\textwidth]{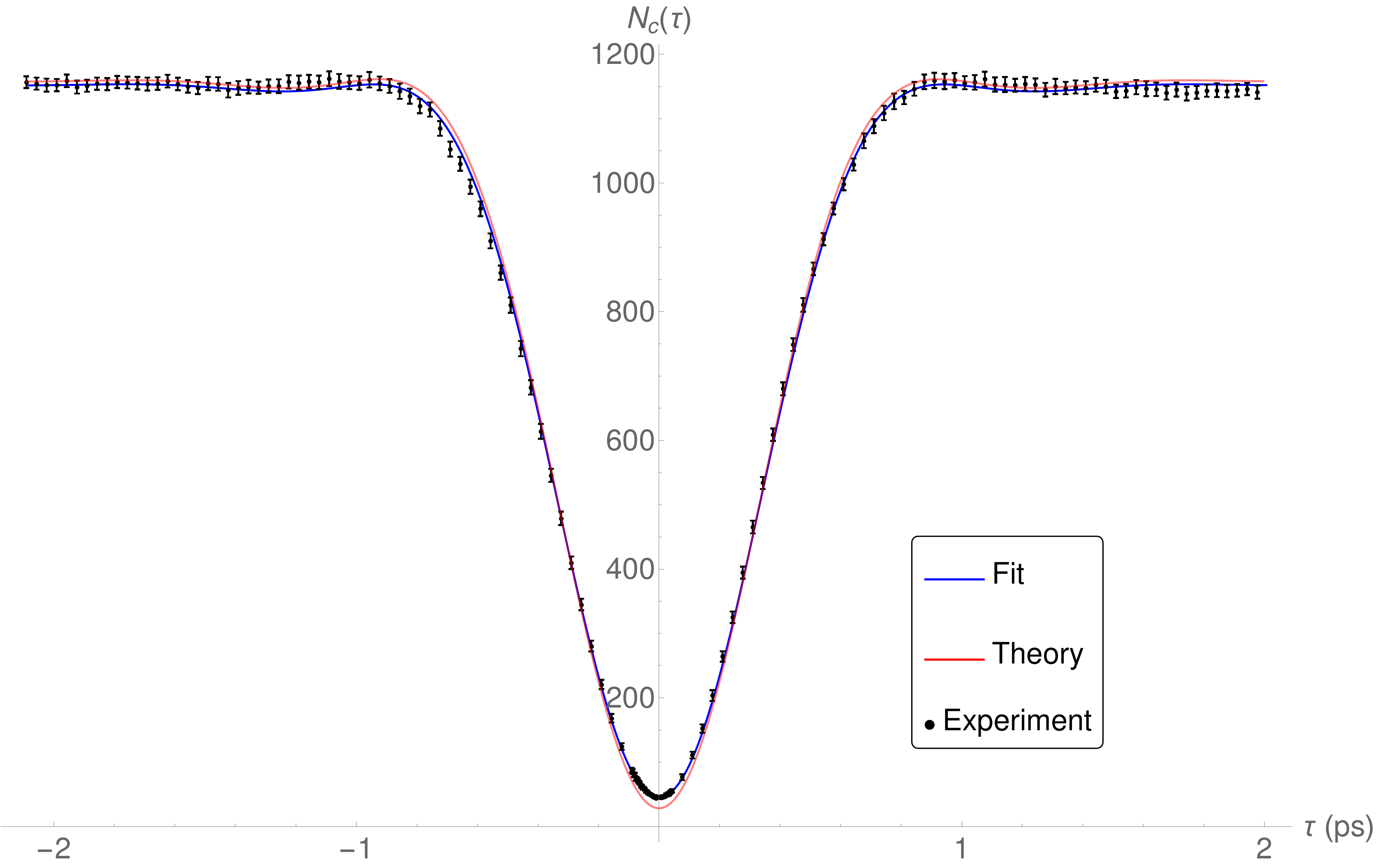}
		\caption{\label{fig:IR}
			Mean coincidence counts measured as a function of time delay between the two input photons are represented by dots. The green dashed line is the theoretical expectation (see Appendix \ref{app:ThEstQ}). The orange solid line is the result of fitting the theoretical result to the data (see Appendix \ref{app:ThFitQ}) resulting in a TPCVD of $96.06\%$. The fitted line overlaps almost completely with the green dashed line (theoretical expectation). The $R$-squared value of the fit is $0.9998$. The error bars represent 95$\%$ confidence intervals for the mean coincidence count at each time delay.
		}
	\end{figure}
	
	\section{Complementarity experiment}
	Classical pulses indeed demonstrate a high degree of anticorrelation indicated by a ``HOM-like'' dip, by applying appropriate relative-phase control between the two inputs.
	As the question of classical-vs-quantum nature is established
	by trying to realize particle-wave complementarity,
	we recombine outputs from the power splitter
	on a second power splitter PS2 to make a Mach-Zehnder interferometer (MZI) to observe interference.
	Whether particlelike or wavelike behavior can be observed in the classical case depends crucially on the relative-phase choice for the inputs.
	Figures~\ref{fig:complementarityC} and \ref{fig:complementarityQ} show the schematics for the classical pulse and photonic versions of this experiment.
	\begin{figure}
		\includegraphics[width=0.48\textwidth]{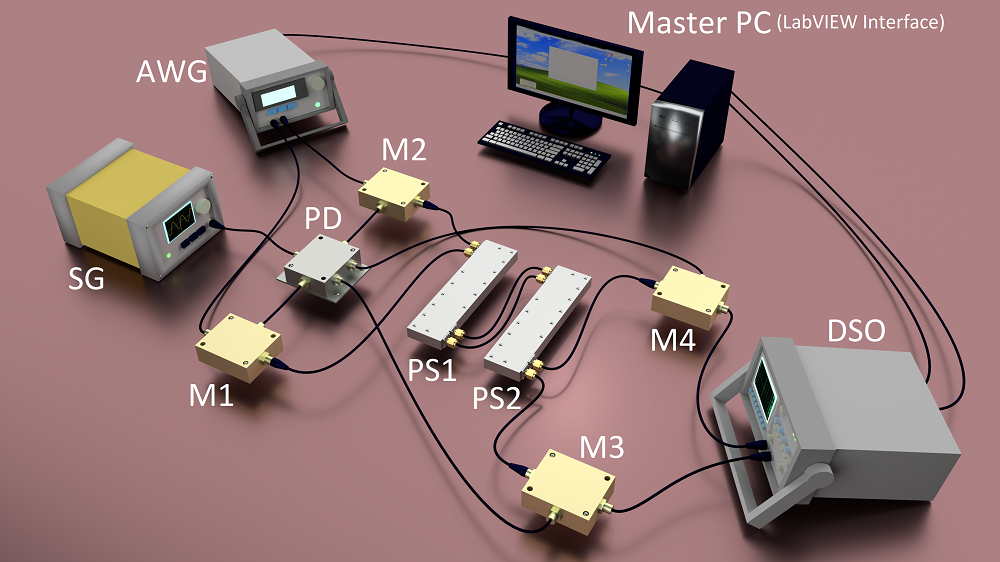}
		\caption{\label{fig:complementarityC}
			The schematic for the complementarity experiment using classical pulses. Detailed description of the setup is in the main text.
		}
	\end{figure}
	
	\begin{figure}
		\includegraphics[width=0.48\textwidth]{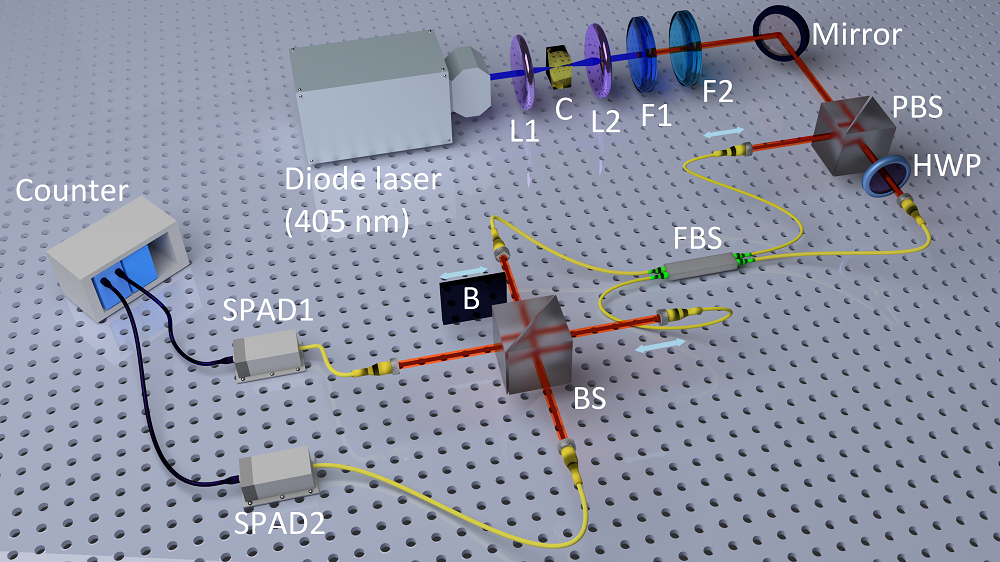}
		\caption{\label{fig:complementarityQ}Schematic for the complementarity experiment using IR photons. Detailed description of the setup is in the main text.
		}
	\end{figure}
	
	Instead of {checking for} complementarity through the presence or absence of interference, we {compare} the classical and quantum cases {using} relevant ratios which only involve measurement at the central maximum of the interference pattern. This is a much simpler test for complementarity as it avoids phase stabilization of the interferometer in the IR domain and changing path lengths in the classical domain which can be non trivial.
	
	We compare the case with both arms of the MZI unblocked (case A) with that in which one of the arms of the MZI is blocked (case B). In case A, the second power splitter or beam splitter recombines the arms to form the original inputs resulting in the probability of coincidence being $1$ in both the classical and the quantum versions. But in case B, the results of the classical and quantum versions differ.
	
	With classical pulses, the output {of PS1} is in either one of the arms of the MZI, and the blocker blocks half of the signals to {PS2}. The unblocked arm simply splits into two resulting in a cross correlation of 50\% compared to that in case {A}.
	
	In the quantum version, the blocker collapses the superposition state to either $\ket{2,0}$ or $\ket{0,2}$ with equal probability, which blocks half of the states, say $\ket{0,2}$, from reaching the second beam splitter. After going through the beam splitter, $\ket{2,0}$ transforms to 
	\begin{align}
	\ket{2,0} \mapsto \frac{1}{\sqrt{2}}\ket{1,1} + \frac{1}{2}\ket{2,0} + \frac{1}{2}\ket{0,2} 
	\end{align}
	yielding a coincidence probability of half. The drop in coincidence probability combined with the fact that half of the superposition states were blocked results in a $25\%$ coincidence rate compared to that in case {A}.
	
	{In experiment}, we complete the MZI after PS1 with another power splitter PS2. The time delay between the two input pulses to PS1 is fixed at $\tau = 0$. The two outputs of PS1 are fed to PS2 {(case A)}. The cross correlation of outputs of PS2 is calculated.
	
	Next we block one arm of the MZI {(case B)}. The connection between the $\Sigma$ port of PS1 and 180$^\circ$ port of PS2 is removed and both ends are terminated with 50-$\Omega$ terminators. The ratio between the cross correlation values {in cases A and B} is computed and found to be $0.4919 \pm 0.0242$. The sources of error are discussed in Appendix \ref{app:error}.
	
	To demonstrate complementarity in the quantum case, we make the two input photons indistinguishable i.e., $\tau=0$ so that the output of the first beam splitter is the superposition of $\ket{2,0}$ and $\ket{0,2}$. The two outputs of the FBS are incident on a 50:50 beam splitter BS through two collimators. One of the collimators {is} placed on a motorized translational stage {to vary the}  path difference ($\Delta l$) between the two arms of the MZI. The output ports of BS are coupled to single-mode fibers that are connected to two single-photon counting modules ($\tau$-SPAD from Pico Quant) for coincidence measurement using a universal quantum device time-tagging unit.
	
	We measure coincidence counts as a function of $\Delta l$ and observe interference. However, to compare with the classical version, we fix $\Delta l=0$, i.e., at the interference maximum. We then compare the coincidence count in case A with that in B. 
	
	To determine the position of the translational stage such that $\Delta l$=0, we make the time delay between the input photons large so that there is no two-photon state in the MZI arms. Then we use the interference pattern between the two arms of MZI to set the stage position for which $\Delta l = 0$. We then make the time delay between input photons zero again. The result of our experiment is a ratio of $0.25 \pm 0.009$. 
	
	{This confirms that in the quantum case, we observe complementarity between $100\%$ TPCVD (particle nature) and interference (wave nature) with the same setting of the sources. The classical signal was not in a superposition state thereby not exhibiting wave nature.}
	
	\section{Conclusion}
	Our theory and experiment elucidate the quantumness of one of the most famous, ubiquitous, and useful quantum optics experiments. In summary, we have shown that the semiclassical theory of photodetection can be applied to calculate the coincidence probability in an HOM experiment, provided the ensemble of inputs is chosen appropriately. The 50\% threshold for the dip in coincidence probability is not the boundary between classical and quantum behaviors of light, but rather a result of the lack of control over the phases of the input pulses. Although classical pulses can be made to switch between exhibiting particlelike and wavelike behaviors by controlling the phase, both cannot manifest with the same choice of phase.  A two-photon state on the other hand shows wave-particle complementarity, by being indivisible in an anticorrelation experiment vs taking a superposition of two paths when interference is measured in an MZI, without any change in the source setting. This wave-particle complementarity is the boundary between the classical and the quantum.
	
	\section{Acknowledgements}
		We thank S. D. Nadella, G. Saha and P. Umesh for their initial technical assistance. S.S. thanks the Canadian Queen Elizabeth II Diamond Jubilee Scholarships program (QES) for funding his visit to the University of Calgary
		during the course of this project. B.C.S. appreciates the VAJRA fellowship support from SERB, Government of India. 

\renewcommand{\appendixname}{APPENDIX}
\appendix
\section{\uppercase{TPCVD from semiclassical theory of photodetection}}
\label{append:TPCVDSemiclassicalTheory}
In classical electromagnetism, light is represented by classical fields. Consider two pulses of linearly polarized light, traveling in perpendicular directions, i.e.,
\begin{align}
\begin{split}
\centering
E_1(\bm{r},t; \bm{k}_1, \omega, 0, 0) = \mathcal{E}(\bm{k}_1 \cdot \bm{r} - \omega t)~  e^{\mathrm{i} (\bm{k}_1 \cdot \bm{r} - \omega t)},
\label{eq:inputFieldOne}
\end{split}\\
\begin{split}
\centering
E_2(\bm{r},t; \bm{k}_2, \omega, \tau, \varphi) = \mathcal{E}(\bm{k}_2 \cdot \bm{r} - \omega t - \omega \tau)~  e^{\mathrm{i} (\bm{k}_2 \cdot \bm{r} - \omega t + \varphi)},
\end{split}
\label{eq:inputFieldTwo}
\end{align}
where $E_i$ is the electric field of the $i^{\text{th}}$ pulse. In general, the field is vectorial, but if both signals have the same polarization, the vector notation can be dropped. $\mathcal{E}$ is the amplitude modulation of the signals (such as Gaussian), $\bm{k}_i$ is the propagation vector, $\omega$ is the angular frequency, $\varphi$ is the relative phase between the two signals and $\tau$ is the time delay between the two pulses. The perpendicularity of the traveling directions of the two pulses necessarily and sufficiently implies that $\bm{k}_1 \cdot \bm{k}_2 = 0$. Furthermore, if the setup is such that the path lengths of both beams are equal, then we ignore the $\bm{r}$ dependence as well. 

In simplified notation, if the two perpendicular pulses serve as inputs to a balanced beam splitter, the outputs are
\begin{equation}
E_\pm(t;\omega, \tau, \varphi) = \frac{1}{\sqrt{2}} \left(E_1(t;\omega, 0, 0) \pm E_2(t; \omega, \tau, \varphi)\right).
\label{eq:outputFieldOne}
\end{equation}
We now introduce fluctuation in the signals by making the phase~$\varphi$ a random variable with a fixed probability distribution $P(\varphi)$ such that
\begin{equation}
\int d\varphi~ P(\varphi)~\cos\varphi = 0,
\label{eq:averageOfCos}
\end{equation}
which yields
\begin{equation}
\left|E_\pm\right|^2 = \frac{1}{2}~ \left(\left|E_1\right|^2 + \left|E_2\right|^2\right)
\label{eq:noSecondOrderInterference}
\end{equation}
thereby revealing that second-order interference between the pulses is not observable.

Time $T$ over which each signal is measured is much larger than input-pulse width. Using Eqs.~\eqref{eq:inputFieldOne}$-$\eqref{eq:outputFieldOne}, with the simplified notation, we obtain
\begin{align}
\int\limits_0^T dt\left|E_\pm(t;\omega,\tau,\varphi)\right|^2 =&\frac{1}{2}\int\limits_0^T dt \left[\mathcal{E}^2(t;\omega) + \mathcal{E}^2(t-\tau;\omega)\right.\nonumber\\
&\left.\pm 2 \mathcal{E}(t;\omega)\mathcal{E}(t-\tau;\omega) \cos\varphi\right].
\label{eq:firstTermOfC}
\end{align}
If $T$ is large compared to the input-pulse width, the first two terms are equal. The third term, which captures the temporal overlap of the two pulses, is a function of time delay and phase. From the definition of cross correlation [Eq.~(\ref{eq:crossCorrelationTheoryExpression}) in main text] and (\ref{eq:firstTermOfC}), we obtain
\begin{align}
C(\tau) &= 1 - \frac{\mathcal{I}^2(\tau;\omega)}{\mathcal{I}^2(0;\omega)}\int d\varphi~ P(\varphi)~ \cos^2\varphi,
\end{align}
where
\begin{equation}
\mathcal{I}(\tau;\omega) = \int\limits_0^T~  dt~ \mathcal{E}(t;\omega)~\mathcal{E}(t-\tau;\omega),
\end{equation}
which yields a TPCVD
\begin{equation}
V = \int  d\varphi~ P(\varphi)\cos^2\varphi,
\label{eq:tpcvdWithPhi}
\end{equation}
which depends on $P(\varphi)$.

\section{\uppercase{Classical description fails complementarity}}\label{app:compTest}
In this appendix we analyze the mathematical details of how the test for complementarity is used to distinguish between the classical and quantum cases. A second beam splitter can be added to the setup to make a Mach-Zehnder interferometer with a phase shifter (creating a phase shift of $\theta$) in one of the arms. 

In the two-photon case, the input state to the second beam splitter is $$\frac{1}{\sqrt{2}}\left(\ket{2,0} -  e^{\text{i}\theta}\ket{0,2}\right)$$.
The output of the second beam splitter is then 
$$\frac{1- e^{\text{i}\theta}}{2\sqrt{2}}\left(\ket{2,0} + \ket{0,2}\right)
+\frac{1+ e^{\text{i}\theta}}{2}\ket{1,1}$$
and therefore, an interference pattern is observed as the phase shifter is rotated. For the trivial case of $\theta=0$ (no phase shift), the output is $\ket{1,1}$ as expected. 

On the other hand, in the case with classical pulses, the phase shifter doesn't give rise to such an interference at the output, as the output of the first beam splitter always travels entirely through one of the arms of the MZI. The outputs fields of the second beam splitter are 
\begin{equation}
\eta_\pm = \frac{1}{\sqrt{2}}\left[E_+ \pm  e^{\text{i}\theta} E_-\right],
\end{equation}
where $\theta$ is the phase introduced in one arm.
Note that when $\varphi \in \{0,\pi\}$ one of $E_\pm$ is always zero and consequently, the terms involving $\theta$ always vanish, making $\left|\eta_\pm\right|^2$ independent of the phase that is introduced by the phase shifter.
In terms of coincidence probability, the phase shifter affects the the coincidence rate in the quantum case but leaves it untouched in the classical case.

Alternatively, if one of the arms is blocked, the coincidence count drops.
Say $\ket{0,2}$ is blocked,
then  a projection operator on the state yields 
\begin{equation}
I\otimes \sum\limits_{n}\ket{0}\bra{n}\left(\frac{\ket{2,0}-\ket{0,2}}{\sqrt{2}}\right) = \frac{1}{\sqrt{2}}\left(\ket{2,0} - \ket{0,0}\right).
\end{equation}
Further, applying the second beam splitter $\hat{B}$ on this state makes the output
\begin{align}
\hat{B}\frac{1}{\sqrt{2}}\left(\ket{2,0} - \ket{0,0}\right) = &\frac{1}{2\sqrt{2}}\left(\ket{2,0} + \ket{0,2} + \sqrt{2}\ket{1,1}\right) \nonumber \\
&- \frac{1}{\sqrt{2}}\ket{0,0}.
\end{align}
This means that the coincidence probability drops from $1$ down to $1/4$.
This drop in probability is a combined effect of the loss of half of the biphotons due to blocking and the absence of interference between the two arms. 

In the classical case the drop in probability of coincidence is different.
If the arm with field $E_-$ is blocked,
\begin{equation}
\eta_\pm=\frac{1}{\sqrt{2}} E_+
\implies\left|\eta_\pm\right|^2 =\frac{1}{2} \left|E_+\right|^2.
\end{equation}
On the other hand, if $E_+$ is blocked,
\begin{equation}
\eta_\pm = \pm\frac{1}{\sqrt{2}} E_- 
\implies~ \left|\eta_\pm\right|^2 = \frac{1}{2} \left|E_-\right|^2.
\end{equation}
In either case, the un-normalized cross correlation function is
\begin{align}
\tilde{C}(\tau) = \frac{I^2 + \mathcal{O}(\tau)}{4},
\end{align}
where 
\begin{equation}
\mathcal{O} (\tau) = \int\limits_{-\infty}^{\infty}  dt ~\mathcal{E}(t) \mathcal{E}(t-\tau) 
\end{equation}
and $I = \mathcal{O}(0)$. Specifically, $C(0) = \frac{I^2}{2}$ compared to $I^2$ when none of the arms are blocked. However, in either case $\left|\eta_+\right|^2 = \left|\eta_-\right|^2$ and hence perfectly correlated, which implies that classical theory predicts that there will never be a click in only one detector. Therefore the classical description predicts that the coincidence probability drops down by a factor of 2 which is different from the prediction of the quantum theory according to which this factor is 4. This result is equivalent to the complementarity test, and separates the classical theory from the quantum one. 

\section{\uppercase{Signal processing in the classical experiment}}\label{append:signalProc}
In the experiment, sinusoidal voltages play the role of the EM radiation. Time delay between input signals $\tau$ and relative phase $\varphi$ are variable parameters. For each value of $\tau$ an ensemble of output signals is generated with values of~$\varphi$ chosen from a chosen probability distribution. The ensemble is used to calculate cross correlation.

Mathematical expressions of the signals generated using an arbitrary waveform generator (AWG) are
\begin{align}
\begin{split}
E_1(t;\omega, \tau, \varphi) = A_1~\mathrm{Exp}\left(-\frac{1}{2}\frac{t^2}{\sigma^2}\right)~ \sin\left(\omega t\right),
\end{split}\\
\begin{split}
E_2(t;\omega,\tau,\varphi) = A_2~ \mathrm{Exp}\left(-\frac{1}{2}\frac{(t-\tau)^2}{\sigma^2}\right)~ \sin\left(\omega t + \varphi \right),
\end{split}
\label{eq: input signals}
\end{align}
where $A_1 = A_2 = 0.05 ~ \mathrm{V}$ are the peaks of the Gaussian envelopes of the two signals. Both signals have Gaussian envelopes with $\sigma = 0.001~\mathrm{s}$, and the sinusoidal wave has a frequency $f = 1~\mathrm{KHz}$ ($\omega = 2\pi f$).

The frequency range in which the power splitter has a splitting ratio of 50:50 is well beyond the maximum frequency that the AWG can produce. So, the effective input is generated by up-converting the frequency of the input signal by using a frequency mixer. The frequency of the local oscillator (LO) signal used for the mixing is $f_L = 1161.999$ MHz. A four-port power divider is used to branch the LO signal into four channels, two of which are used to up-convert the input signals, and the other two are used to down-convert the output of the power splitter. After the power splitter splits the inputs, the outputs need to be down-converted for measurement. 

The measurement device used is an oscilloscope, two of whose channels are used to measure the two outputs of the power splitter.
As the oscilloscope has an upper limit to the frequencies that it can measure, the high-frequency output of the power splitter is down-converted using the LO. The result of down-conversion is a mixture of low and high frequencies. The oscilloscope acts as a low-pass filter and measures only the low-frequency components, making the effective output
\begin{equation}
E_\pm(t;\omega, \tau,\varphi) =\frac{A_L^2}{2} \frac{E_1(t;\omega,\tau,\varphi) \pm E_2(t;\omega,\tau,\varphi)}{\sqrt{2}}.
\end{equation}
Although the output depends on $A_L$, the cross correlation function does not depend on it because of normalization.
\section{\uppercase{Data acquisition and post-processing in the classical experiment}}\label{append:dataAcPostProc}
The digital oscilloscope samples the voltage signal at a sampling rate of two giga-samples per second and consequently the recorded signal is a time series of voltages $E_+^{(n)}(\tau,\varphi)$ and $E_-^{(n)}(\tau,\varphi)$. Each time series contains a thousand points which implies that the time interval between successive points is $\Delta t = \frac{1}{2\times 10^9}$ s and the total acquisition time is $T = \frac{1000}{2 \times 10^9}$ s.

To calculate $C(\tau)$ we take the ensemble average over the fluctuating $\varphi$. We focus our attention to fluctuation that is ergodic, described by a time-independent probability distribution $P(\varphi)$. The fluctuation can then be simulated by a discrete random process in which a pair of signals is generated with phase difference $\varphi$, chosen from the said distribution.

\section{\uppercase{Minimum $N$ for good statistics}}
\label{sec: number of samples}
The relative phase ($\varphi$) of the input signals is averaged over so that there is no second-order interference as we are interested in studying only the fourth-order interference between the inputs. Consequently, the average integrated intensities of the output in both arms of the beam splitter (the normalization factor for the cross correlation) are independent of $\tau$.
Therefore, to ensure absence of the second-order interference, we determine the minimum number of samples (of $\varphi$) required to estimate the constant average-integrated-intensity within 5\% error with a confidence level of 95\%
using the two-tailed test \cite{dekking2005modern}. 

Let
\begin{equation}
\mu(\omega,\tau) = \int d\varphi~ P(\varphi) \int\limits_0^T  dt~ \left|E_\pm(t;\omega,\tau,\varphi)\right|^2
\end{equation}
be the true average integrated intensity of the output signals and 
\begin{equation}
\mu^*(\omega,\tau) = \frac{1}{N} \sum\limits_{i=1}^{N} \int\limits_0^T  dt~ \left|E_\pm(t;\omega,\tau,\varphi_i)\right|^2
\end{equation}
be the average calculated from the sample outputs, where $N$ is the number of samples recorded. Let
\begin{equation}
t = \sqrt{N} \frac{\mu^*(\omega,\tau) - \mu(\omega,\tau)}{\sigma(\omega,\tau)},
\end{equation}
where $\sigma(\omega,\tau)$ is the true standard deviation of the integrated intensities for each $\tau$. From the central limit theorem we know that for large enough $N$, the distribution of $t$ will be closely approximated by the Normal distribution. As the 95\% confidence interval for a Normal distribution is $\left(-1.96,1.96\right)$, the 95\% confidence interval (C.I.) for $\mu^*(\omega,\tau)$ is
\begin{equation}
\text{C.I.} = \left(\mu(\omega,\tau)-\frac{1.96\sigma(\omega,\tau)}{\sqrt{N}},\mu(\omega,\tau)+\frac{1.96\sigma(\omega,\tau)}{\sqrt{N}}\right)
\end{equation}
If we want the error in the estimation of $\mu$ to be $\pm 5\%$ with 95\% confidence, the width of the C.I. must be 10\% of $\mu$;
i.e.,
\begin{equation}
2 \times\frac{1.96~\sigma(\omega,\tau)}{\sqrt{N}} = 0.1\mu(\omega,\tau).
\label{eq:confIntervalError}
\end{equation}
Equation~(\ref{eq:confIntervalError}) is then solved to get the minimum number of samples $N$ required,
\begin{equation}
N(\omega,\tau) = \left(\frac{1.96\sigma(\omega,\tau)}{0.05\mu(\omega,\tau)}\right)^2.
\end{equation}

In the 100\% TPCVD case, the phase between the input signals is chosen uniformly from the set $\{0, \pi\}$, so the probability distribution of $\varphi$ is
\begin{equation}
P(\varphi) = \frac{1}{2}~ \left(\delta(\varphi)+\delta(\varphi-\pi)\right)
\end{equation}
Figure \ref{fig:minN100} shows $N_{\text{min}}$ as a function of $\tau$.
\begin{figure}[H]
	\centering
	\includegraphics[width=0.49\textwidth]{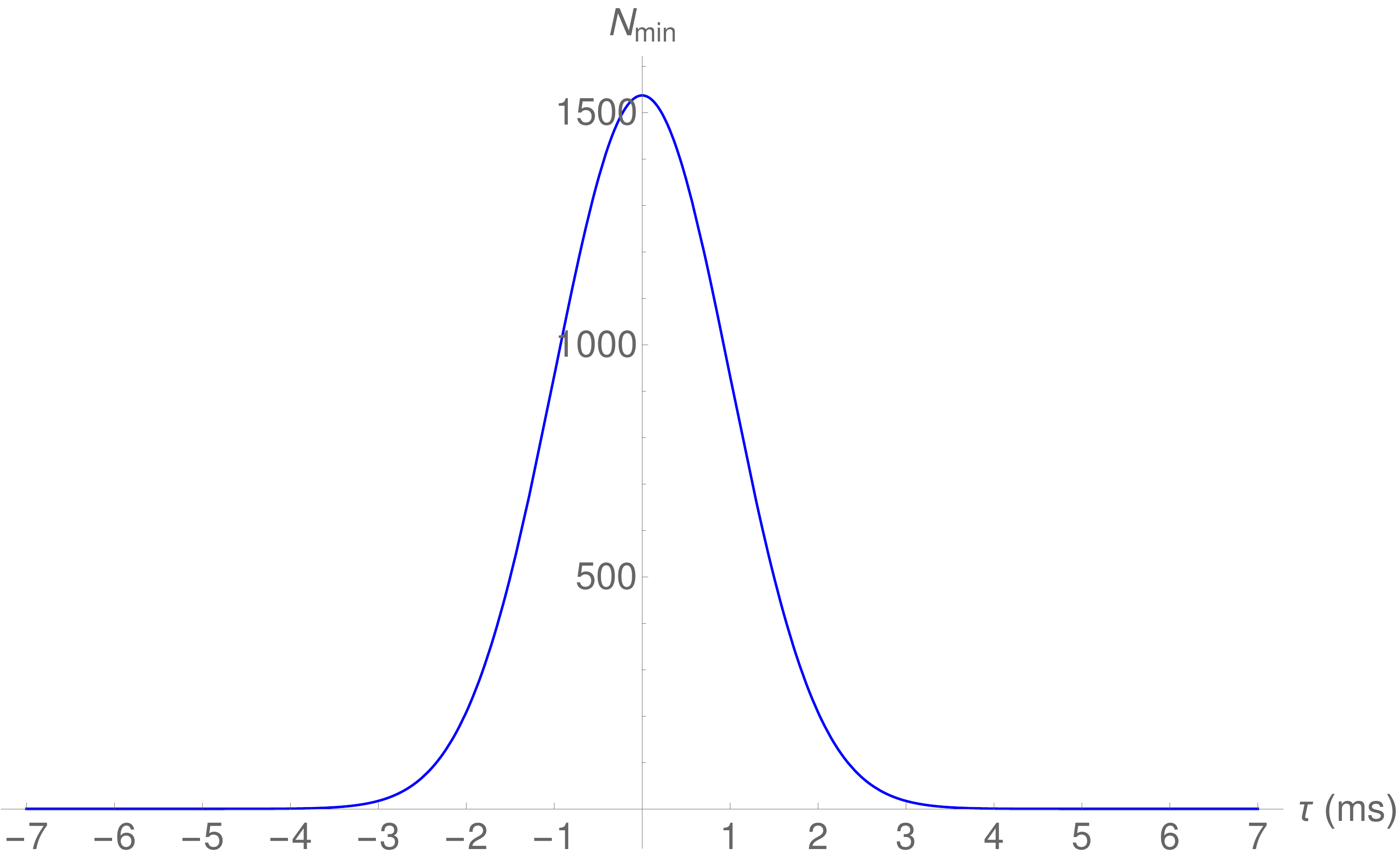}
	\caption{Plot of minimum number of samples (of $\varphi$) $N_{\text{min}}$ vs the time delay $\tau$ between the input signals, for the experiment demonstrating 100\% TPCVD.}
	\label{fig:minN100}
\end{figure}
The 50\% TPCVD experiment is done in two versions. First the phase is chosen uniformly from the continuous interval $[0,2\pi)$ for which
\begin{equation}
P(\varphi) = \frac{1}{2\pi}.
\end{equation}
In the other version, the phase is chosen uniformly from the set $\left\{0,\frac{\pi}{2}, \pi,\frac{3\pi}{2}\right\}$ with probability distribution
\begin{equation}
P(\varphi) = \frac{1}{4}~ \left(\delta(\varphi)+\delta\left(\varphi-\frac{\pi}{2}\right)+\delta(\varphi-\pi)+\delta\left(\varphi-\frac{3\pi}{2}\right)\right).
\end{equation}
In either case, the means and standard deviations for all values of $\tau$ are identical and hence $N_{\text{min}}$ is given by the graph in Fig.~\ref{fig:minN50}.
\begin{figure}[H]
	\centering
	\includegraphics[width=0.49\textwidth]{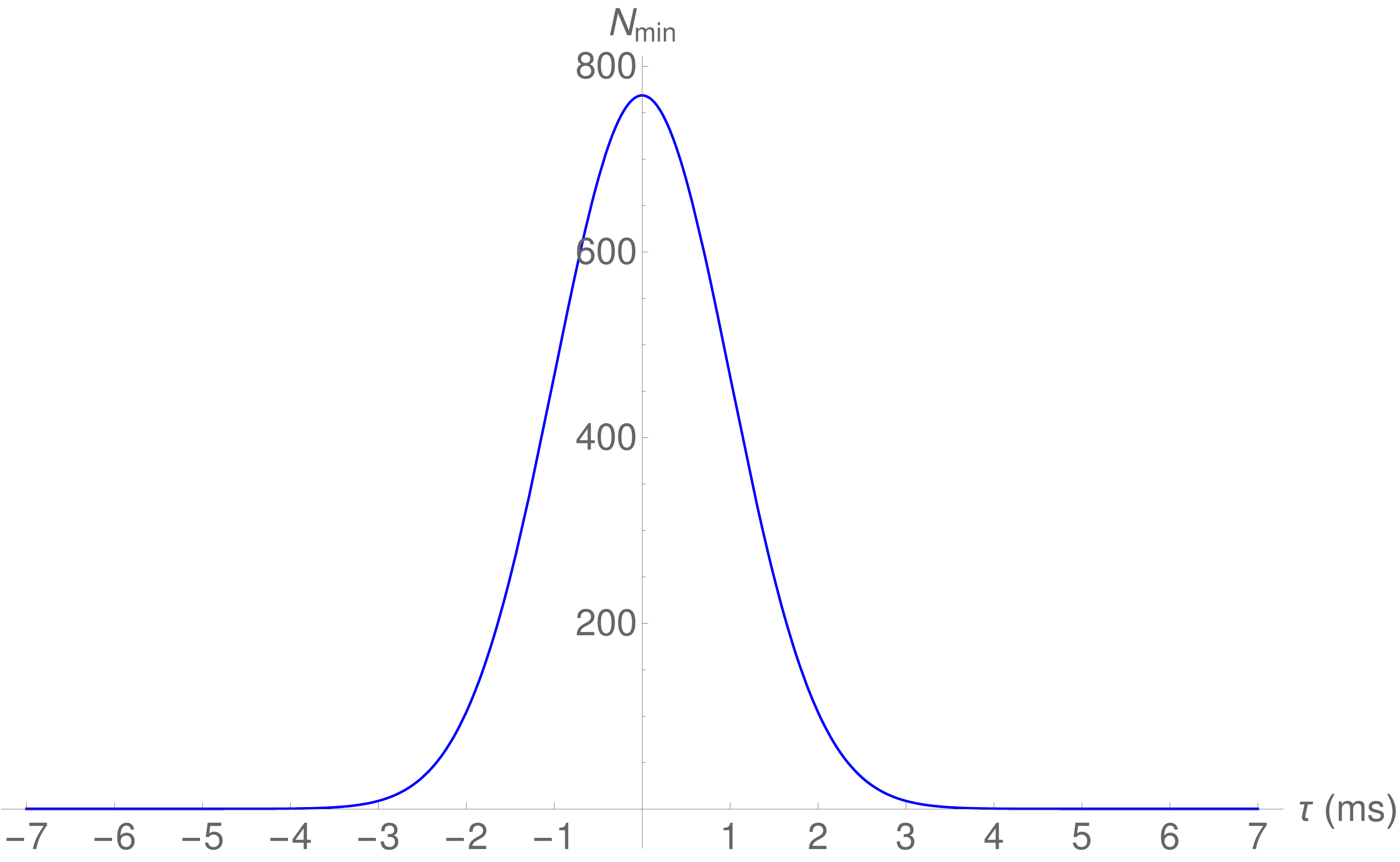}
	\caption{Plot of minimum number of samples (of $\varphi$) $N_{\text{min}}$ vs the time delay $\tau$ between the input signals, for the experiment demonstrating 50\% TPCVD.}
	\label{fig:minN50}
\end{figure}
The number of samples to be generated in the experiment is guided by the graphs in Figs.~\ref{fig:minN100} and \ref{fig:minN50}.

\section{\uppercase{The cross correlation graph for 50\% with continuous variation of phase}}\label{append:50continuous}
The main text presents the cross correlation graph for the case in which the relative phase between the input signals is varied over the discrete set $\{0,\pi/2,3\pi/2,\pi\}$. The same 50\% dip in visibility is also seen when the phase is varied continuously over the interval $[0,2\pi)$ as shown in Fig.~\ref{fig:50continuousPhase}.
\begin{figure}[H]
	\centering
	\includegraphics[width=0.48\textwidth]{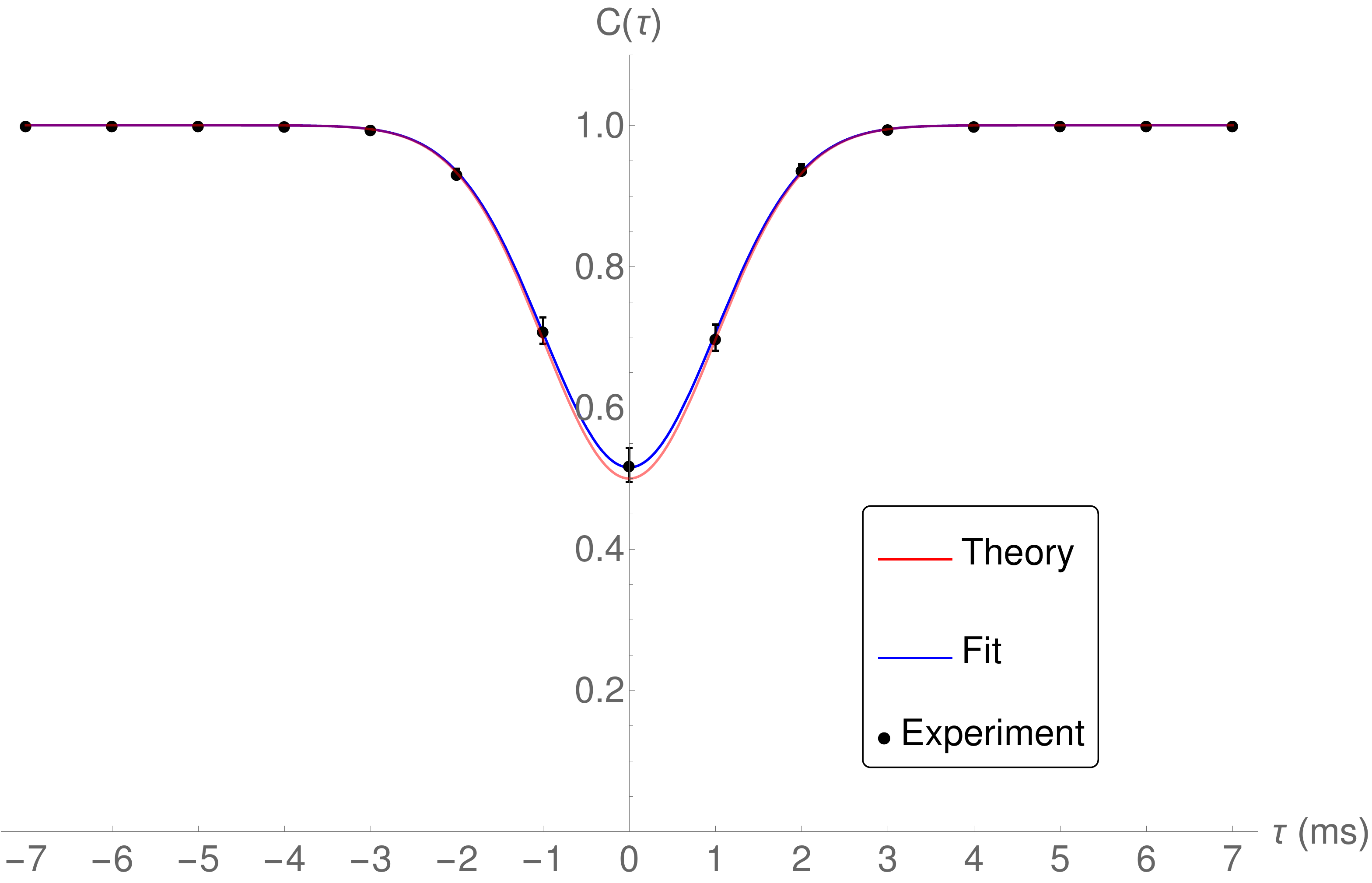}
	\caption{The normalized cross correlation of the output pulses, as a function of the time delay $\tau$ between the input signals when the phase between them is uniformly random over the continuous interval $[0,2\pi)$. The experimental result is represented by the dots. The green dashed ``Theory'' curve is the result of the theoretical calculation and shows a 50\% dip. The attenuation of the signals causes an amplitude mismatch which results in a dip slightly less than 50\%. Taking the amplitude mismatch as a fit parameter, the cross correlation is fitted to the experimental data to get the orange solid ``Fit'' curve.}
	\label{fig:50continuousPhase}
\end{figure}

\section{\uppercase{Confidence intervals for photon counts using bootstrap}}\label{append:bootstrapPhoton}
The result of the experiment is a graph of coincidence counts vs actuator position (which is proportional to the time delay between the input photons). One hundred iterations of the experiment were done, which resulted in a sample of 100 values of the coincidence counts for each actuator position. Consider the sample of coincidence counts $X=\{c_1, c_2, \cdots, c_{100}\}$ at a particular actuator position, say $l$. The estimate of the counts is then the mean over $X$ which is the sample mean, i.e., $\bar{c}$. The confidence interval for $\bar{c}$ was then found by employing a statistical bootstrap. The method involves creating a large number of sets by resampling with replacement from the original set. The size of the resampled set is the same as that of the original. Let one such resampled set be $X^*$ and its mean $\bar{c}^*$. We define a quantity
\begin{equation}
\delta = \bar{c}^* - \bar{c}.
\end{equation}
Repeating the above step a large number of times, in our case 10 000, we get a sample for $\delta$. To find a 95 percentile confidence interval, we pick the $2.5^{\text{th}}$ and the $97.5^{\text{th}}$ percentile of this sample, $\delta_{0.025}$ and $\delta_{0.975}$. The confidence interval for $\bar{c}$ is then simply $[\delta_{0.025} + \bar{c}, \delta_{0.975} + \bar{c}]$. We then repeat the bootstrap method for all values of the actuator positions. These confidence intervals were plotted as the error bars for the mean value at each actuator position.

\section{\uppercase{Sources of error}}\label{app:error}
\subsection{Classical experiment}
The dominant source of error was the amplitude mismatch of the outputs, which is a result of the each signal undergoing unequal attenuation as they pass through different paths of the circuit. 
The remaining sources of error have negligible contribution, however, for completeness, they must be discussed.

Every instrument used in the circuit has an uncertainty in the measurement of the respective quantity. From the data sheets of each instrument provided by the manufacturer the instrumental errors in frequency, time delay, voltage, frequency of the local oscillator and amplitude of the local oscillator were used to find the maximum error due to instrumentation. The error in the beam-splitting ratio of the power splitters was found from their characterization. The relative error is plotted in Fig.~\ref{fig: errors from uncertainties}.

Other sources of errors include finite sampling rates of the AWG and the oscilloscope. However, the rates are sufficiently high for near perfect sampling of the signals.

\begin{figure}[H]
	\centering
	\includegraphics[width=0.49\textwidth]{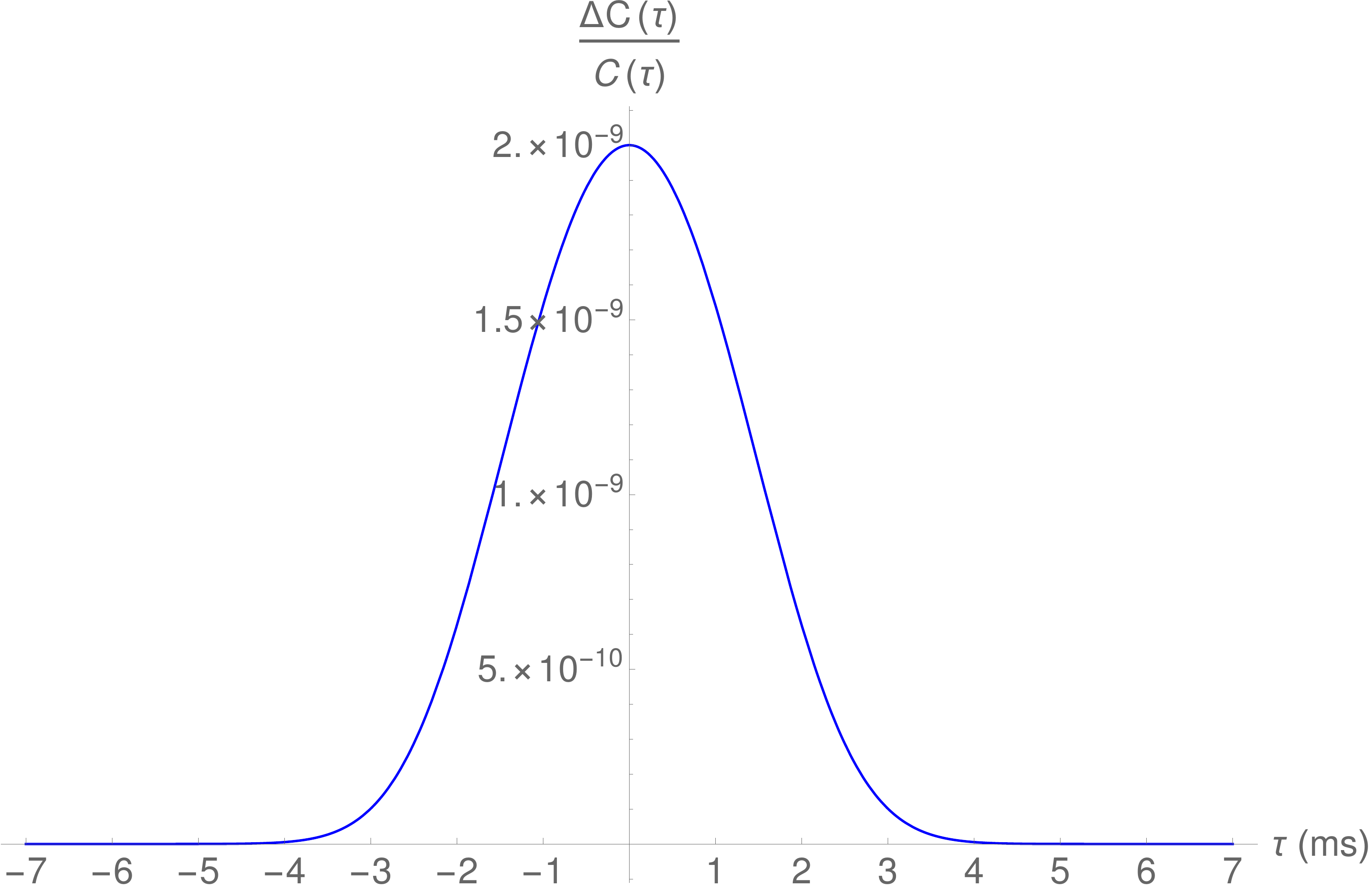}
	\caption{The relative error in the cross correlation of the output signals due to various instrumental errors, as a function of the time delay $\tau$ between the input signals.}
	\label{fig: errors from uncertainties}
\end{figure}

\subsection{Quantum experiment}
Using the quantum description, the coincidence probability at the output of a (lossless) beam splitter is given by the well-known result by Hong-Ou-Mandel,
\begin{align}
C(\delta\tau)=&\left|T\right|^4+\left|R\right|^4 \nonumber \\
&-2\left|T\right|^2 \left|R\right|^2\int d\tau~{\text{Re}\{g(\tau+\delta\tau)g^{*}(\tau-\delta\tau)\}},
\label{eq:quantumCoincidence}
\end{align}
where $g(\tau)$ is the Fourier transform of $f(\omega)$; $f(\omega)$ is the normalized joint spectral amplitude (JSA) of the two input photons, $\omega$ being the difference in the frequency components of the two photons [$\omega=(\omega_{2}-\omega_{1})/2$]. The time delay between the two input photons to the beam splitter is  $\delta\tau$. If the two photons are completely indistinguishable, then a balanced beam splitter causes the coincidence probability to drop to zero, i.e., $C(0)=0$. $C(\delta\tau)=1/2$ for $\delta\tau\gg\tau_{c}$, where $\tau_{c}$ is the coherence time of the single photons. 

In practice, however, different components may introduce some distinguishability in different degrees of freedom (frequency, polarization, spatial mode, etc.) of the two input photons. Let $\eta$ be the probability that the input photons retain indistinguishability and $\zeta$ the probability both photons fall on the same port of the beam splitter. Then
\begin{widetext}
	\begin{equation}
	C(\delta\tau)=(1-\zeta)\left[|T|^4+|R|^4-2|T|^2 |R|^2\eta\int d\tau~{\text{Re}\{g(\tau+\delta\tau)g^{*}(\tau-\delta\tau)\}}\right]+2\zeta |T|^2 |R|^2
	\label{eq: fitModel}
	\end{equation}
\end{widetext}
which is the expression used to fit the coincidence probability curve to the experimental data.

\section{\uppercase{Theoretical estimate for HOM profile for the quantum experiment}}\label{app:ThEstQ}
We have plotted coincidence counts vs time delay, taking into account various instrumental errors [Eq.~(\ref{eq: fitModel})].
The dominant contributions to these errors are from the non ideal extinction ratio of the polarizing beam splitter (PBS), imperfect rotation of the half waveplates and the effect of the interference filter. An ideal PBS transmits only horizontally ($\ket{H}$) polarized light and reflects vertically ($\ket{V}$) polarized light. But in practice this may not happen owing to the imperfect extinction ratios. Similarly, errors in the half waveplate rotation may introduce distinguishability in polarization. The transmittance of the band-pass filter affects the JSA of the input photons. 

The values of these error parameters were obtained from the datasheets of the respective instruments. The PBS has extinction ratios 1000:1 in the transmission arm and, 52:1 in the reflection arm. So, $T_{H}/T_{V}=1000$ and $R_{V}/R_{H}=52$, where $T_{H}, R_{H}$ represents transmission and reflection probabilities of $\ket{H}$ polarized light through the PBS. $T_{V}$ and $R_{V}$ are the transmission and reflection probabilities of $\ket{V}$ polarized light. The probability of both $\ket{H}$ and $\ket{V}$ photons being in any one of the output arms of the PBS [$\zeta$ in Eq.~(\ref{eq: fitModel})] is 
\begin{equation}
\zeta=T_{H}T_{V}+R_{H}R_{V}=0.0201.
\end{equation}
The transmission and reflection coefficients of the fiber beam splitter (FBS) are $\left|T\right|^{2}=0.52, \left|R\right|^{2}=0.48$. The half wave-plate (HWP) rotation should be such that the angle ($\theta$) of the HWP axis with respect to the horizontal axis is $45^{\text{o}}$, in order to make both the photons possess the same polarization. But the HWP rotation has an error of $\pm 1^{\text{o}}$, leading to 
\begin{equation}
\eta=\sin^{2}2\theta=0.9988.
\end{equation}
The transmission-vs-frequency data $F(\omega)$ for the filter are provided by the manufacturer. With the filter, the joint-spectral amplitude of the two input photons becomes
\begin{equation}
f(\omega) = F(\omega)\phi(\omega),
\end{equation}
where $\phi(\omega)$ is the joint-spectral amplitude of the photons without the filter. We have assumed $\phi(\omega)$ to be a Gaussian distribution with standard deviation $\sigma$.

In order to compare with the experimental data, we have multiplied a scaling factor ($K$) to the coincidence probability $C(\delta\tau)$ [coincidence counts=$K\times C(\delta\tau)$]. We put values to all relevant parameters, i.e., $\eta$, $\zeta$, $T$, $R$, and fit the function   $K\times C(\delta\tau)$ with the experimental data, while taking only $K, \sigma$ as fit parameters. We get a fit (shown as ``Theory'' in Fig.~5 in the main text) with $R$-squared $0.9998$ for $\sigma=0.59$ nm, $K=2301$, resulting in an expected TPCVD of 97.56$\%$.

\section{\uppercase{Finding a fit to the experimental result}}\label{app:ThFitQ}
Although the major systematic errors have been taken into account above, there are additional errors that are untraceable, like dispersion and rotation of polarization as the photon passes through different fibers and components, spatial mode mismatch in the two fiber inputs of the FBS, etc. Consequently, the experimental result has a slight deviation from the theoretical estimate. To get a better fit to the data, we use the expression $K\times C(\delta\tau)$ [see Eq.~(\ref{eq: fitModel})] to find a fit with $\eta$, $\zeta$, $K$, $\sigma$ as fit-parameters.

The ``Fit'' line in Fig.~5 is a result of such a fit. The best fit, with $R$-squared $0.9998$ was achieved for $\sigma=0.581$ nm, $K=2303$, $\zeta=0.038$, $\eta=0.9995$, resulting in a TPCVD of $96.06\%$.

	\bibliography{HOM}

\end{document}